\documentclass[aps,amsmath,amssymb,superscriptaddress,twocolumn]{revtex4-2}

\usepackage{amsmath}
\usepackage{graphicx}
\usepackage[dvipsnames]{color}
\usepackage{hyperref}
\usepackage{psfrag}
\usepackage{stmaryrd}
\usepackage{amssymb}
\usepackage{wasysym}
\usepackage{float}
\usepackage{color}

\usepackage[T1]{fontenc}
\usepackage[english]{babel}
\usepackage{lmodern}
\usepackage{graphicx} 
\usepackage{times}
\usepackage{amsmath,amssymb}
\usepackage{color}

\begin{document}
\def\bnablaperp#1{\boldsymbol{\nabla}_\perp ^{#1}}


\title{Rigorous ultimate scaling in rapidly rotating steady convection}

\author{Gabriel Hadjerci}
\email{hadjerci.gabriel.pv4@osaka-u.ac.jp}
\author{Shingo Motoki}%
\author{Genta Kawahara}
 
\affiliation{%
Graduate School of Engineering Science, University of Osaka. 1-3 Machikaneyama, Toyonaka, Osaka 560–8531, Japan.
}%

\begin{abstract}
Rapidly rotating Rayleigh–Bénard convection admits a class of exact steady single-mode solutions describing high-amplitude convection cells. Using a matched asymptotic analysis in the high-Rayleigh-number limit, we obtain a rigorous characterization of their bulk and boundary-layer structure, yielding explicit scaling laws for the Nusselt and Reynolds numbers, including their dependence on the horizontal wavenumber. We show that, for suitable wavenumbers, these solutions attain the diffusivity-free ultimate scalings frequently assumed for geophysical and astrophysical convection, with additional enhancing logarithmic corrections. This reveals a specific mechanism through which rapidly rotating convection can approach ultimate heat transport via coherent columnar structures with well-defined horizontal scales.
\end{abstract}

\maketitle

\textit{Introduction--}Thermal convection under global rotation is a fundamental process in astrophysical and geophysical systems, where it drives large-scale flows relevant to climate dynamics and magnetic-field generation~\cite{spiegel_convection_1971,stevenson_turbulent_1979,marshall_openocean_1999,aurnou_rotating_2015,vasil_rotation_2021}. A canonical framework for its study is rotating Rayleigh–Bénard convection (RRBC), in which a fluid layer subject to vertical gravity and uniform rotation is confined between two horizontal plates maintained at a fixed temperature difference~\cite{chandrasekhar_hydrodynamic_1961,rossby_study_1969,king_heat_2012,ecke_turbulent_2023}. In the rapid-rotation limit, relevant to many natural systems, the non-hydrostatic quasigeostrophic (NHQG) equations provide an accurate asymptotic description of RRBC~\cite{julien_new_1998,julien_reduced_2007,julien_heat_2012,julien_nonlinear_2016,julien_rescaled_2025,maffei_inverse_2021,sprague_numerical_2006}.  These reduced equations allow a distinctive class of exact steady solutions known as single-mode solutions (SMS). Beyond serving as useful benchmarks for numerical simulations, these solutions describe coherent columnar structures with various horizontal patterns such as rolls, squares, and hexagons ~\cite{bassom_strongly_1994,julien_fully_1999,sprague_numerical_2006,grooms_asymptotic_2015}.  Originally identified in studies of the full Boussinesq equations in the rapid-rotation asymptotic limit~\cite{bassom_strongly_1994} (see~\cite{matthews_asymptotic_1999} for its magnetoconvection counterpart), they have attracted considerable interest because they constitute analytical representations of high-amplitude convection cells, thereby extending weakly nonlinear theory into the strongly nonlinear regime.

In addition to rapid rotation, geophysical and astrophysical systems often operate under extreme thermal forcing, producing strongly turbulent flows that preclude a complete analytical treatment. Consequently, scaling laws for key macroscopic quantities, such as global heat flux, are typically inferred from phenomenological arguments or dimensional analysis~\cite{kraichnan_turbulent_1962,spiegel_convection_1971,stevenson_turbulent_1979,marshall_openocean_1999,grossmann_scaling_2000,aurnou_connections_2020}. {An alternative approach, however, is to study invariant solutions. Indeed, although not fully turbulent, they can capture essential mechanisms or establish constraints on turbulent scaling behavior~\cite{hassanzadeh_wall_2014,motoki_maximal_2018,motoki_optimal_2018,motoki_multi-scale_2021,kooloth_coherent_2021,wen_steady_2022,deguchi_high-taylor-number_2023,he_high-rayleigh-number_2026}. In addition, these solutions maintain analytical accessibility, thereby facilitating comprehensive characterization.} In this spirit, Grooms~\cite{grooms_asymptotic_2015} carried out a detailed analysis of SMS in the high-Rayleigh-number limit and established exact inequalities for the heat flux and other quantities. Guided by these bounds, we perform a matched asymptotic analysis that yields a rigorous characterization of SMS in both the bulk and boundary layers. This approach allows us to go beyond bounds and derive explicit scaling laws for heat and momentum transport through the rigorous expression of the solution, thereby converting previously known inequalities into precise asymptotic relations.

\begin{figure*}
    \centering
    \includegraphics[width=0.98\linewidth]{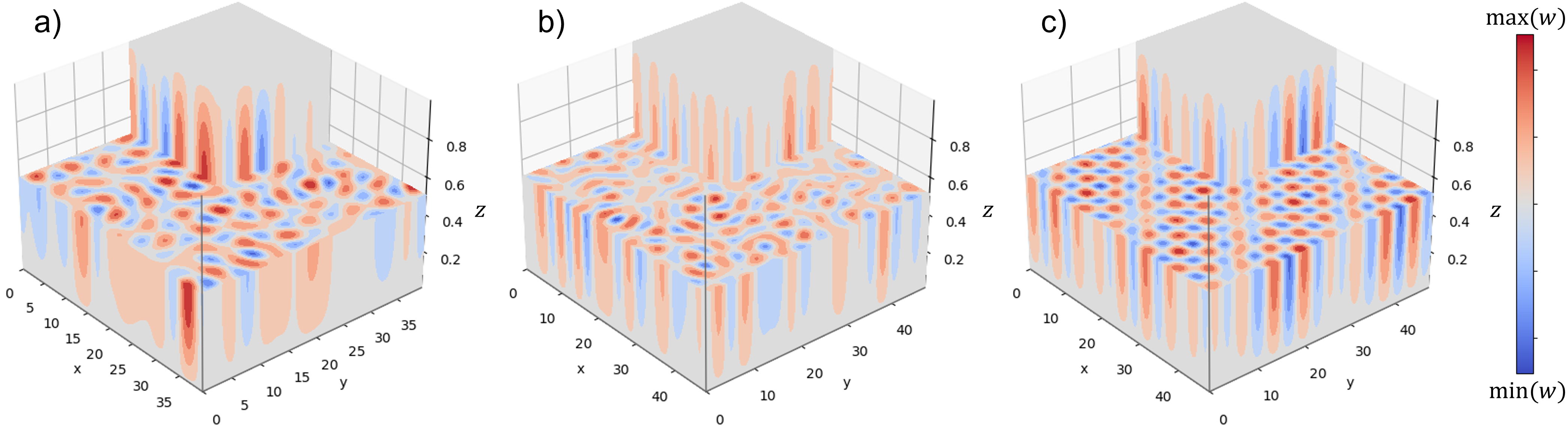}
    \caption{Three-dimensional visualizations of the instantaneaous vertical velocity field at $\widetilde{Ra}=10$ and $Pr=1$ for: a) DNS of full Boussinesq RRBC with $E=10^{-6}$; b) DNS of the NHQG equations; c) SMS with $k=1.3048$ found using a Newton-Krylov method.}
    \label{fig:velocity}
\end{figure*}

\textit{NHQG equations--}Under the Boussinesq approximation~\cite{spiegel_boussinesq_1960}, RRBC is governed by three dimensionless parameters: the Rayleigh, Prandtl, and Ekman numbers, defined as:
\begin{equation}
    Ra = \frac{\alpha g \Delta T H^3}{\nu\kappa}; \qquad Pr = \frac{\nu}{\kappa}; \qquad {E} = \frac{\nu}{2\Omega H^2}\, ,
\end{equation}
where $H$ is the depth of the layer, $g$ the gravity acceleration, $\Delta T$ the temperature difference, $\Omega$ the rotation rate, $\alpha$ the thermal expansion coefficient of the fluid, $\nu$ its kinematic viscosity, and $\kappa$ its thermal diffusivity. 

The NHQG equations are derived through a double-scale asymptotic expansion with $\varepsilon=E^{1/3}$ as the small parameter~\cite{sprague_numerical_2006}. The horizontal coordinates $(x,y)$ vary on the reduced scale $\varepsilon H$,  while the vertical coordinate $Z$ remains scaled by $H$. At leading order, the dimensionless velocity field takes the form $(u,v,w)=(\partial_y\psi,-\partial_x\psi,-\bnablaperp{2}\phi)$, where $\psi$ and $\phi$ are the toroidal and poloidal potentials and where $\boldsymbol{\nabla}_\perp$ denotes the horizontal gradient. The dimensionless temperature field is decomposed into a mean and a fluctuation component, such that $\theta = \overline{\theta}+ \varepsilon\, \theta^\prime$, where $\overline{\cdot}$ denotes a horizontal and temporal average. All fields are periodic in the horizontal directions. At the boundaries $Z=0,1$, we impose impermeability, $\phi=0$, and fixed temperature, $\theta^\prime=0$ with $\overline{\theta}(0)=1$ and $\overline{\theta}(1)=0$. These conditions imply additional constraints through the governing equations. In particular, stress-free conditions $\partial_Z\psi=0$ at $Z=0,1$ follow from impermeability and are not imposed independently.

In the NHQG system, the Nusselt number, quantifying the efficiency of the convective heat flux, is given by:
\begin{equation}
    Nu  = 1+Pr\,\langle w\theta^\prime \rangle\, ,
\end{equation}
where $\langle\cdot\rangle$ denotes a volume and time average in a statistically stationary state. Due to the NHQG scaling for velocities, the Reynolds number naturally takes the reduced form:
\begin{equation}
    \widetilde{Re}= Re E^{1/3}=\sqrt{\langle u^2 \rangle+\langle v^2 \rangle+\langle w^2 \rangle}\, ,
\end{equation}
where $Re$ is the usual Reynolds number based on $H$ and the dimensional root-mean-square velocity. In contrast to the non-reduced equations, the NHQG system depends only on $Pr$ and on the reduced Rayleigh number $\widetilde{Ra} = RaE^{4/3}$. Consequently, both $Nu$ and $\widetilde{Re}$ depend only on these two parameters. A central issue in strongly forced convection is whether the flow reaches a diffusivity-free regime, consistent with turbulence theory~\cite{kraichnan_turbulent_1962,spiegel_convection_1971,stevenson_turbulent_1979,marshall_openocean_1999}. This regime—often referred to as geostrophic turbulence or the ultimate regime in the literature~\cite{barker_theory_2014,bouillaut_experimental_2021,song_scaling_2024,hadjerci_rapidly_2024,hadjerci_radiatively_2024,kannan_beyond_2025}—is characterized, in the rapidly rotating limit, by the scaling relations:
\begin{equation}\label{eq:ultimate_scaling}
Nu\sim\widetilde{Ra}^{3/2}Pr^{-1/2} \hspace{3mm} \text{ and } \hspace{3mm} \widetilde{Re}\sim\widetilde{Ra}Pr^{-1} \, .
\end{equation}

\textit{Single-mode solutions--}The NHQG equations admit a distinguished class of exact steady solutions of separable form:
\begin{equation}
\left(\begin{array}{c}
\psi \\
\phi \\
\theta^\prime
\end{array}\right) = h(x,y)\,\left(\begin{array}{c}
\widehat{\psi}(Z) \\
\widehat{\phi}(Z) \\
\widehat{\theta^\prime}(Z)
\end{array}\right)\, ,
\end{equation}
where $h$ satisfies the planform equation $\nabla_\perp^2 h = -k^2 h$ with $k>0$ and $\overline{h}=1$~\cite{julien_fully_1999,sprague_numerical_2006}. For fixed $\widetilde{Ra}$, a steady solution of this form exists if and only if:
\begin{equation}\label{eq:singlemode}
d_{ZZ}\phi_* + \left(\frac{\widetilde{Ra}Nu\,k^2}{1+\phi_*^2}-k^6\right)\phi_*=0\, ,
\end{equation}
where $\phi_*(Z)=k Pr\, \widehat{\phi}(Z)$ and $\phi_*(0)=\phi_*(1)=0$. The Nusselt number is given by:
\begin{equation}\label{eq:Nusselt}
Nu=\left( \int_0^1\frac{1}{1+\phi_*^2}\,dZ \right)^{-1}\, .
\end{equation}
This equation also owns an integrated form, obtained by multiplying~(\ref{eq:singlemode}) by $d_Z \phi_*$ and integrating:
\begin{equation}\label{eq:integrated}
\left(d_Z\phi_*\right)^2=P_0^2+k^6\phi_*^2-\widetilde{Ra}Nu\, k^2 \log\left(1+\phi_*^2\right)\, ,
\end{equation}
with $P_0=d_Z \phi_*(0)$~\cite{bassom_strongly_1994}. These nonlinear steady solutions are referred to as single-mode, as their horizontal structure involves a single wavenumber $k$. Once $\phi_*$ is determined, all remaining fields follow from:
\begin{equation}
Pr\,\hat\psi  =\frac{1}{k^3}d_Z\phi_*\, ,\ \hat\theta^\prime  =\frac{1}{k} \frac{Nu\, \phi_*}{1+\phi_*^2}\, ,\ \partial_Z\overline{\theta} =-\frac{Nu}{1+\phi_*^2}\, .
\end{equation}

Using the rescaled potential $\phi_*$ instead of $\widehat{\phi}$ removes the explicit dependence on the Prandtl number from both the SMS equation~(\ref{eq:singlemode}) and the Nusselt number~(\ref{eq:Nusselt}). The resulting description thus depends only on the planform wavenumber $k$ and the reduced Rayleigh number $\widetilde{Ra}$. To solve~(\ref{eq:singlemode}), it is convenient to introduce the flux-based reduced Rayleigh number $\widetilde{Ra}_P=\widetilde{Ra}Nu$ and treat $\widetilde{Ra}$ as an output rather than a control parameter. The latter is recovered \textit{a posteriori} using the Nusselt number~(\ref{eq:Nusselt}). Due to its nonlinearity, Eq~(\ref{eq:singlemode}) may allow multiple solutions for fixed $(k,\widetilde{Ra}_P)$. Besides the conductive state $\phi_*=0$, nonlinear branches arise with a structure analogous to vibrating-string modes. The fundamental solution is of fixed sign, symmetric about $Z=1/2$, with a single interior extremum, whereas higher harmonics alternate in sign and correspond to physically unrealistic velocity and temperature profiles. In this study, we thus restrict our attention to the fundamental solution with $\phi_*\geq 0$ —without loss of generality since the equation is invariant under $\phi_*\to-\phi_*$. This solution is computed numerically for a given pair $(k,\widetilde{Ra}_P)$ using a shooting method that enforces Dirichlet boundary conditions at $Z=0,1$. For comparison, we also perform direct numerical simulations (DNS) using the pseudo-spectral solver CORAL~\cite{miquel_coral_2021} for both the NHQG equations and the full Boussinesq equations. Three-dimensional visualizations of the instantaneous vertical velocity at $\widetilde{Ra} = 10$ are shown in Fig.~\ref{fig:velocity}, alongside the corresponding field from a SMS at the same $\widetilde{Ra}$. At this moderate $\widetilde{Ra}$, where turbulence is not yet fully developed, all three systems exhibit similar columnar structures.

\begin{figure*}
    \centering
    \includegraphics[width=0.99\linewidth]{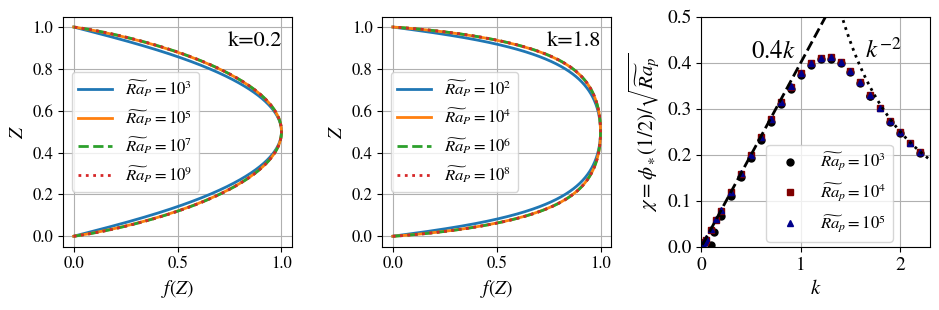}
    \caption{Normalized solution $f(Z)=\phi_*(Z)/\phi_*(1/2)$ with increasing $\widetilde{Ra}$, for $k=0.2$ (left) and  $k=1.8$ (middle). The right panel shows $\chi=\phi_*(1/2)/\sqrt{\widetilde{Ra}_P}$ as a function of $k$ for various $\widetilde{Ra}_P$.}
    \label{fig:normalized_profiles}
\end{figure*}

\textit{Bulk structure--}A closer inspection of~(\ref{eq:singlemode}) reveals that the main complexity arises from the nonlinear term ${\widetilde{Ra}_P\,k^2\phi_*}/{(1+\phi_*^2)}$. Expanding this term is thus the natural route toward simplifying the equation. To assess the validity of such an expansion, it is necessary to compare $\phi_*^2$ with unity. Since the solution reaches its maximum at mid-depth, the Nusselt number expression~(\ref{eq:Nusselt}), yields: $Nu < 1 + \phi_*^2(1/2)$. In the asymptotic limit $\widetilde{Ra} \gg 1$, $Nu$ becomes large~\cite{grooms_asymptotic_2015}, so that $\phi_*^2(1/2) \gg 1$. By contrast, the boundary conditions impose $\phi_*=0$ at the bounding surfaces, implying $\phi_* \ll 1$ in their vicinity. Consequently, any matched asymptotic analysis of the SMS must account for at least two distinct regions: a bulk region where $\phi_*^2 \gg 1$, and boundary layers where $\phi_*^2 \ll 1$.

According to Grooms, nontrivial solutions of~(\ref{eq:singlemode}) may exist only if:
\begin{equation}\label{eq:Range_k}
 k<\widetilde{Ra}^{1/4} \hspace{5mm} \text{and} \hspace{5Mm} 2\pi \widetilde{Ra}\geq \pi^2/k^2+k^4. 
\end{equation}
Within this admissible range, we first focus on the mid-plane maximum. A first characterization follows from the bounds established by Grooms:
\begin{equation}\label{eq:A_bounds}
\frac{\widetilde{Ra}_P}{k^4+8k^{-2}} -1 \leq \left[\phi_*(1/2)\right]^2< \frac{\widetilde{Ra}_P}{k^4} -1\, .
\end{equation}
For fixed $k$, these bounds imply $\phi_*(1/2) \sim \sqrt{\widetilde{Ra}_P}$  at large $\widetilde{Ra}_P$. Accordingly, there exists a function of $k$, $\chi(k)$, such that $\phi_*(1/2) \approx \sqrt{\widetilde{Ra}_P} \chi(k)$. Since $\phi_* \gg 1$ in the bulk region, the SMS equation thus reduces to:
\begin{equation}\label{eq:bulk_equation}
d_{ZZ}f + \frac{k^2}{\chi^2(k)} f^{-1}-k^6 f=0\, ,
\end{equation}
where $f(Z)=\phi_*(Z)/\phi_*(1/2)$ is the normalized solution. By construction, $f(1/2)=1$, while symmetry about the mid-plane gives $d_Zf(1/2)=0$. Since~(\ref{eq:bulk_equation}) equipped with these boundary conditions is independent of $\widetilde{Ra}_P$, the normalized solution is likewise independent of $\widetilde{Ra}_P$. Thus, for fixed $k$, the bulk solutions tend toward a self-similar profile as $\widetilde{Ra}_P$ increases. Fig.~\ref{fig:normalized_profiles} confirms this behavior. For fixed $k$, the normalized solutions collapse onto a common curve at large $\widetilde{Ra}_P$. The collapse extends to higher derivatives in the bulk, but breaks down near the boundaries, as $d_Z f(0)=P_0/\sqrt{\widetilde{Ra}}$ increases with $\widetilde{Ra}_P$ (see Eq.~(\ref{eq:P0})).

Simple approximations for $\chi(k)$ can be derived in the large and small-$k$ limits. First, in the large-$\widetilde{Ra}_P$ limit, the bounds~(\ref{eq:A_bounds}) give $(k^4+8k^{-2})^{-1}\le\chi^2<k^{-4}$, implying:
\begin{equation}\label{eq:chi_large_k}
\chi(k) \approx k^{-2}\, \text{ for } k\gg1\, .
\end{equation}
Next, assuming $\chi(k)$ is finite, it comes that $k^6 \ll k^2/\chi^2$ for $k\ll1$. The $k^6$ term in the bulk equation~(\ref{eq:bulk_equation}) thus become negligible. As will be discussed in the next section, it is also negligible in the boundary layer equation. Thus, for $k\ll1$, the full SMS equation~(\ref{eq:singlemode}) reduces to:
\begin{equation}
d_{ZZ}\phi_* + \frac{\widetilde{Ra}_P k^2 \phi_*}{1+\phi_*^2}=0\, ,\hspace{1mm} \text{with }\ \phi_*(0)=\phi_*(1)=0\, .
\end{equation}
This equation depends only on the combined parameter $\widetilde{Ra}_P k^2$, so does the amplitude $\phi_*(1/2)$. Consistency thus requires:
\begin{equation}\label{eq:chi_small_k}
\chi(k)\approx a k\, \text{ for } k\ll 1\, ,
\end{equation}
where $a$ is a constant. The bounds~(\ref{eq:A_bounds}) imply $a\ge 8^{-1/2}\approx 0.3536$~, while numerically we find $a\approx0.4$. The two reference lines in the right panel of Fig.~\ref{fig:normalized_profiles} confirm the asymptotic behavior predicted by Eqs.~(\ref{eq:chi_large_k}) and~(\ref{eq:chi_small_k}). For Eq.~(\ref{eq:chi_small_k}), deviations at very small $k$ arise from the growth of the critical value $\widetilde{Ra}_c = \pi^2/k^2 + k^6$~\cite{chandrasekhar_hydrodynamic_1961} and from the breakdown of the asymptotics near onset. Inversely, its domain of validity broadens as $\widetilde{Ra}_P$ increases.

In summary, in the limit $\widetilde{Ra}\gg1$, the bulk solution takes the form:
\begin{equation}
\phi_*(Z)=\chi(k) \sqrt{\widetilde{Ra}_P}f(Z)
\end{equation}
where $f$ satisfies~(\ref{eq:bulk_equation}) and is independent of $\widetilde{Ra}_P$, while $\chi(k)\approx k^{-2}$ for $k\gg1$, and $\chi(k) \approx a k$ for $k\ll 1$.

\textit{Boundary layer solution--} In the boundary layer, where $\phi_* \ll 1$, the nonlinear term in~(\ref{eq:singlemode}) reduces to $\widetilde{Ra}_P k^2 \phi_*$. Moreover, using~(\ref{eq:Range_k}) together with the assumption $Nu \gg 1$, we obtain $k^6<\widetilde{Ra}k^2\ll\widetilde{Ra}_P k^2$. It follows that the $k^6$ term is negligible in the boundary-layer equations. The equation in the lower boundary layer (the upper layer follows by symmetry) thus becomes:
\begin{equation}
d_{ZZ}\phi_* + \widetilde{Ra}_P\,k^2\phi_*=0\, ,\hspace{2mm} \text{with }\left\{\begin{array}{l}
\phi_*(0)=0\\
d_Z\phi_*(0)=P_0
\end{array}\right. .
\end{equation}
This linear equation can be solved exactly, and the solution is:
\begin{equation}
\phi_*(Z)=\frac{P_0}{k\sqrt{\widetilde{Ra}_P}}\sin\left( k\sqrt{\widetilde{Ra}_P}\, Z \right)\, .
\end{equation}
This sinusoidal shape prevents convergence of the boundary-layer solution to the bulk solution, compromising the existence of a consistent matching condition. One could introduce a buffer layer to ensure a proper connection. However, we rather exploit the integrated form~(\ref{eq:integrated}), which offers a natural link between the two regions. Evaluating~(\ref{eq:integrated}) at $Z=1/2$ yields:
\begin{equation}\label{eq:P0}
P_0= k\sqrt{\widetilde{Ra}_P}\times \sqrt{\log\left(1+\widetilde{Ra}_P \chi^2\right)-k^4\chi^2}\, .
\end{equation}
Hence, for $\widetilde{Ra}\gg1$, the boundary-layer solution becomes:
\begin{equation}\label{eq:BL_solution}
\phi_*(Z)=\sqrt{\log\left(1+\widetilde{Ra}_P \chi^2\right)-k^4\chi^2}\,\sin\left( k\sqrt{\widetilde{Ra}_P}\, Z \right).
\end{equation}

\begin{figure}[h!]
    \centering
    \includegraphics[width=0.99\linewidth]{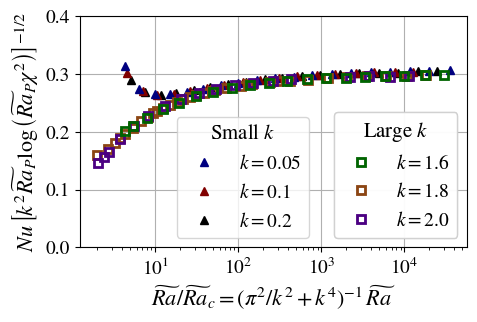}
    \caption{Nusselt number of the SMS compensated by the prediction of the scaling law~(\ref{eq:Nusselt_scaling}) as a function of the reduced Rayleigh number normalized by its value at the onset of convection. For small wavenumbers $k=(0.05, \, 0.1,\, 0.2)$, we use the approximation $\chi(k)=0.4 k$, whereas for large wavenumbers $k=(1.6,\, 1.8, \, 2)$, we take $\chi(k)=k^{-2}$.}
    \label{fig:scalings}
\end{figure}

\textit{Scaling laws--} To estimate the Nusselt number, we split the integral~(\ref{eq:Nusselt}) into bulk and boundary-layer contributions:
\begin{equation}
1-\frac{1}{Nu}= \int_{\delta_{BL}}^{1-\delta_{BL}}\frac{\phi_*^2}{1+\phi_*^2}\, dZ + 2 \int_{0}^{\delta_{BL}} \frac{\phi_*^2}{1+\phi_*^2}\, dZ  , 
\end{equation}
where $\delta_{BL}$ denotes the boundary layer thickness. Using $\phi_*\gg1$ in the bulk and $\phi_*\ll1$ in the boundary layers yields:
\begin{equation}
Nu\approx\frac{1}{2\delta_{BL}}\, .
\end{equation}
The boundary-layer solution~(\ref{eq:BL_solution}) remains valid as long as $\phi_*\ll1$. We therefore identify $\delta_{BL}$ as the height at which $\phi_* = 1$. Solving~(\ref{eq:BL_solution}) for $\phi_* = 1$ then provides an estimate of $\delta_{BL}$:
\begin{equation}
\delta_{BL}\approx\frac{\arcsin \left[\left(\log\left( 1+\widetilde{Ra}_P\, \chi^2  \right)- k^4\chi^2\right)^{-1/2}\right]}{k\sqrt{\widetilde{Ra}_P}}\, .
\end{equation}  
In the asymptotic limit $\widetilde{Ra}\gg 1$, the bounds on the wavenumber~(\ref{eq:Range_k}) imply:
$\log\left( 1+\widetilde{Ra}_P\, \chi^2 \right)- k^4\chi^2 \gg 1$. Hence, using the corresponding expansion of $\arcsin$ yields:
\begin{equation}\label{eq:Nusselt_scaling}
Nu\sim \left[k^2 \widetilde{Ra}_P \, \log\left(\widetilde{Ra}_P\, \chi^2\right)\right]^{1/2} \, .
\end{equation}
This scaling relation is fully consistent with Grooms’ bounds. Fig.~\ref{fig:scalings} shows the Nusselt number compensated by the prediction~(\ref{eq:Nusselt_scaling}) as a function of the reduced Rayleigh number normalized by its critical value—this normalization collapses the curves at fixed $k$ and improves readability without affecting the scaling. For small wavenumbers $k=(0.05,0.1,0.2)$ we use $\chi(k)=0.4 k$, while for $k=(1.6,1.8,2)$ we take $\chi(k)=k^{-2}$. At large $\widetilde{Ra}$, all curves reach a common plateau, confirming that~(\ref{eq:Nusselt_scaling}) captures the asymptotic behavior of the Nusselt number. 

\begin{figure}[h!]
    \centering
    \includegraphics[width=0.99\linewidth]{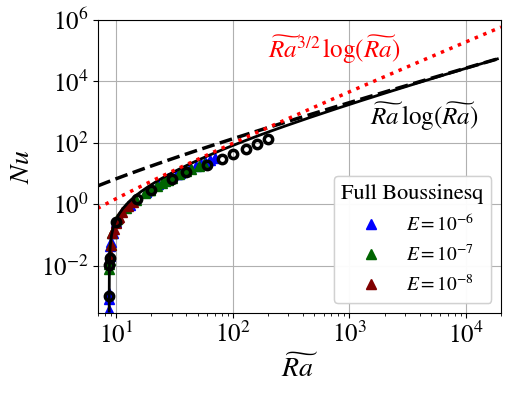}
    \caption{Nusselt number as a function of the reduced Rayleigh number. Black circles: DNS of the NHQG equations. Triangles: DNS of the full Boussinesq equations. Black solid line: SMS with $k=1.3048$. Black dashed line: SMS asymptotic scaling for constant $k$~(\ref{eq:fixed_k_scaling}). Red dotted line: SMS scaling for $k\sim\widetilde{Ra}^{1/4}Pr^{-1/4}$~(\ref{eq:Nusselt_opti}).}
    \label{fig:Nusselt}
\end{figure}

For fixed $k$, we recover the scaling: $Nu\sim\widetilde{Ra}\, \log(\widetilde{Ra}Nu)$ previously discussed in the literature~\cite{bassom_strongly_1994,julien_fully_1999}. Using the large-$x$ asymptotic properties of the Lambert $W$ function, solution of $W(x)e^{W(x)}=x$~\cite{corless_lambert_1996}, this scaling becomes:
\begin{equation}\label{eq:fixed_k_scaling}
Nu\sim\widetilde{Ra}\, \log\left(\widetilde{Ra}\right)\, .
\end{equation}
This scaling is verified in Fig.~\ref{fig:Nusselt}, where $Nu$ for SMS at the critical wavenumber $k_c \approx 1.3048$ is shown as a function of $\widetilde{Ra}$. We also plot DNS results for comparison. The Nusselt number in DNS exhibits a similar behavior to that of SMS with $k \approx k_c$, but remains slightly lower. This difference can be attributed to the fact that pure columnar structures transport heat more efficiently, acting as straight conduits between the two plates.

A scaling for the Reynolds number can be derived from the kinetic energy budget. In the NHQG equations~\cite{sprague_numerical_2006}, it is written:
\begin{equation}
\frac{d\, \langle E_c\rangle_V}{dt} = \frac{\widetilde{Ra}\,\left\langle w\theta^\prime\right\rangle_V}{Pr} - \Big\langle(\bnablaperp{2}\psi)^2+(\bnablaperp{}w)^2\Big\rangle_V\, ,
\label{eq:energy_buget_ARM}
\end{equation}
where $E_c=[(\bnablaperp{}\psi)^2 + (\bnablaperp{2}\phi)^2]/2$ and $\langle \cdot\rangle_V$ denotes a volume average. Since SMS are steady solutions, (\ref{eq:energy_buget_ARM}) yields:
\begin{equation}\label{eq:ReNu_relation}
k^2\,\widetilde{Re}^2=\frac{\widetilde{Ra}\,(Nu-1)}{Pr^2}\, .
\end{equation}
This equality directly involves the scaling law for the Reynolds number:
\begin{equation}\label{eq:Reynolds}
\widetilde{Re}\sim\frac{\sqrt{\widetilde{Ra} Nu}}{Pr\, k}\, ,
\end{equation}
which becomes, at fixed $k$:
\begin{equation}
\widetilde{Re}\sim\widetilde{Ra} Pr^{-1} \left[\log\left(\widetilde{Ra}\right)\right]^{1/2} .
\end{equation}

\textit{Ultimate scaling--}In contrast to previous results~\cite{bassom_strongly_1994,julien_fully_1999}, our scaling laws for the Nusselt and Reynolds numbers~(\ref{eq:Nusselt_scaling},\ref{eq:Reynolds}) retain an explicit dependence on the wavenumber $k$. This enables us to assess whether single-mode solutions can reach the ultimate scaling regime~(\ref{eq:ultimate_scaling}) through an appropriate choice of $k$. The key to determining this is the exact relation~(\ref{eq:ReNu_relation}), which shows that the Nusselt and Reynolds numbers are intrinsically coupled. In particular, it establishes that $Nu$ and $\widetilde{Re}$ can reach the ultimate scaling simultaneously only if the wavenumber satisfies:
\begin{equation}\label{eq:optimal_k}
k \sim \widetilde{Ra}^{1/4}Pr^{-1/4}\, .
\end{equation}
Notably, the $\widetilde{Ra}^{1/4}$ scaling coincides with the exponent that produces the largest bounds for $Nu$ identified by Grooms. In the regime $Pr \ll \widetilde{Ra}$, typical of many astrophysical systems~\cite{van_kan_bridging_2025}, the above condition implies $\chi(k)\approx k^{-2}$. Substituting this expression into~(\ref{eq:Nusselt_scaling},\ref{eq:Reynolds}) and using the asymptotic properties of the Lambert $W$ function yields:
\begin{align}
Nu &\sim\widetilde{Ra}^{3/2}Pr^{-1/2} \log(\widetilde{Ra})\, \label{eq:Nusselt_opti} ,\\
\widetilde{Re}& \sim \widetilde{Ra}Pr^{-1} \left[\log\left(\widetilde{Ra}\right)\right]^{1/2}\, .\label{eq:Reynolds_opti}
\end{align}
These relations correspond to the ultimate scaling predictions, supplemented by logarithmic factors on the right-hand side. In contrast to the more common situation in which logarithmic corrections appear on the left-hand side and reduce heat transport~\cite{kraichnan_turbulent_1962}, here they enhance it. Consequently, coherent columnar structures described by SMS, can approach the ultimate regime with a slightly steeper scaling, via a suitable contraction of their horizontal scale. The scaling~(\ref{eq:Nusselt_opti}) is shown in red in Fig.~\ref{fig:Nusselt}. The available DNS data do not allow one to determine the effective scaling at very large $\widetilde{Ra}$. However, our results suggest that sustaining ultimate scaling at higher $\widetilde{Ra}$ would require a shrinking of turbulent plumes and other columnar structures involved in convective heat transport.

\textit{Discussions--}We analyzed SMS of the NHQG equations using a matched asymptotic analysis in the limit $\widetilde{Ra}\gg1$. This yields a rigorous characterization of the boundary layers and shows that, for fixed $k$, the bulk approaches a self-similar profile governed by a reduced equation independent of $\widetilde{Ra}_P$. These results provide rare analytical access to the high-$\widetilde{Ra}$ convection regime. In particular, we were able to derive rigorous scaling laws for the Nusselt and Reynolds numbers, including their dependence on the horizontal wavenumber $k$. These results go beyond Grooms’ foundational bounds by establishing precise asymptotic relations, revealing how coherent columnar structures approach the ultimate scaling regime through a $\widetilde{Ra}$-dependent contraction of their horizontal scale.

Recent DNS of rapidly rotating Rayleigh–Bénard convection with stress-free boundaries~\cite{kannan_beyond_2025} report a regime in which the Nusselt number reaches ultimate scaling while the Reynolds number deviates from its predicted behavior. Within the SMS framework, the exact scaling relation~(\ref{eq:Reynolds}) provides a natural interpretation to this observation. Indeed, unless $k$ satisfies~(\ref{eq:optimal_k}), ultimate scaling cannot hold simultaneously for both quantities. In particular, if $Nu$ follows the ultimate scaling at fixed $k$, one instead obtains:
\begin{equation}
\widetilde{Re} \sim \left(\widetilde{Ra}/Pr\right)^{5/4}\, ,
\end{equation}
which is remarkably close to DNS observations (Ref.~\cite{kannan_beyond_2025} suggests the exponent $11/8$). This agreement supports the idea that SMS can capture key ingredients of the scaling behavior in the fully turbulent regime, motivating further exploration of their connection to the chaotic state.\\

\textit{Acknowledgements--}This work was supported by the Japan Science and Technology Agency (JST), PRESTO (JPMJPR23OC) and by JSPS KAKENHI Grant Numbers JP23K2267, JP24KF0239, JP25K01157, and JP25K07567. G. Hadjerci acknowledges financial support as a Japan Society for the Promotion of Science International Research Fellow. This research used the supercomputer system of Academic Center for Computing and Media Studies, Kyoto University. The authors would like to thank J. Jimenez for helpful discussions at the 2025 Madrid Turbulence Workshop, which improved their understanding of the single-mode equation.

\bibliography{ASMS}

\end{document}